\documentclass[letter, traditabstract]{aa}
\usepackage{graphicx}
\usepackage{xspace}
\usepackage{natbib}
\usepackage{longtable}
\usepackage{lscape}
\usepackage{txfonts}
\def\um{$\mu$m\xspace}

\def\12c{$^{12}$C\xspace}
\def\h2{H$_2$\xspace}
\def\h2o{H$_2$O\xspace}
\def\so2{SO$_2$\xspace}
\def\co2{CO$_2$\xspace}
\def\co{CO\xspace}

\def\c2h2{C$_2$H$_2$\xspace}
\def\sio{SiO\xspace}
\def\sis{SiS\xspace}
\def\tio2{TiO$_2$\xspace}
\def\tio{TiO\xspace}

\begin{document}

   \title{When an old star smolders:}
   \subtitle{On the detection of hydrocarbon emission from S-type AGB stars}
   \author{K. Smolders 
          \inst{1} \fnmsep \thanks{Aspirant Fellow of the Fund for Scientific Research, Flanders}
          \and
          B.~Acke\inst{1}\fnmsep \thanks{Postdoctoral Fellows of the Fund for Scientific Research, Flanders}
          \and
          T.~Verhoelst\inst{1}\fnmsep $^{\star\star}$
          \and
          J.A.D.L.~Blommaert\inst{1}
          \and
          L.~Decin\inst{1,4} \fnmsep $^{\star\star}$
          \and
          S.~Hony\inst{2}
          \and
          G.~C.~Sloan\inst{3}
          \and
          P.~Neyskens\inst{5} \fnmsep \thanks{Fellowship ``boursier F.R.I.A'', Belgium}
          \and
          S.~Van Eck\inst{5}
          \and
          A.~A.~Zijlstra\inst{6}
          \and
          H.~Van Winckel\inst{1}
          }

   \offprints{K. Smolders}

   \institute{Instituut voor Sterrenkunde (IvS),
              Katholieke Universiteit Leuven,
              Celestijnenlaan 200 D,
              B-3001 Leuven, Belgium
         \and
             Service d'Astrophysique,
             CEA Saclay,
             Bat.709 Orme des Merisiers,
             91191 Gif-sur-Yvette, France
         \and
             Cornell University,
             Astronomy Department,
             Ithaca, NY 14853, USA
         \and
             Universiteit van Amsterdam, Sterrenkundig Instituut ``Anton
             Pannekoek", Kruislaan 403, 1098 SJ Amsterdam, The Netherlands
         \and
             Institut d'Astronomie et d'Astrophysique (IAA), Universit\'e 
             Libre de Bruxelles, C.P.226, Boulevard du Triomphe,
             B-1050 Bruxelles, Belgium
         \and
             Jodrell Bank Centre for Astrophysics, Alan Turing Building, 
             University of Manchester, Oxford Road, Manchester, M13 9PL, UK}  
 \date{\today}

\abstract{Polycyclic aromatic hydrocarbons (PAHs) produce characteristic infrared emission bands that have been observed in a wide range of astrophysical environments, where carbonaceous material is subjected to ultraviolet (UV) radiation. Although PAHs are expected to form in carbon-rich AGB stars, they have up to now only been observed in binary systems where a hot companion provides a hard radiation field. In this letter, we present low-resolution infrared spectra of four S-type AGB stars, selected from a sample of 90 S-type AGB stars observed with the infrared spectrograph aboard the Spitzer satellite. The spectra of these four stars show the typical infrared features of PAH molecules. We confirm the correlation between the temperature of the central star and the centroid wavelength of the 7.9~\um feature, present in a wide variety of stars spanning a temperature range from 3\,000 to 12\,000~K. Three of four sources presented in this paper extend this relation towards lower temperatures. We argue that the mixture of hydrocarbons we see in these S-stars has a rich aliphatic component. The fourth star, BZ~CMa, deviates from this correlation. Based on the similarity with the evolved binary TU~Tau, we predict that BZ~CMa has a hot companion as well.}


\maketitle

\section{Introduction}
Polycyclic aromatic hydrocarbon (PAH) molecules are large molecules containing carbon and hydrogen atoms. They carry typical infrared (IR) emission features that have been observed in many astrophysical environments \citep{Tielens2008}. Since these features are generally attributed to the IR fluorescence of ultraviolet-pumped molecules, we expect PAH features to arise from regions where carbon-rich material is exposed to ultraviolet (UV) radiation \citep{Leger1984, Cohen1985, Puget1989, Allamandola1989}. PAHs play a major role in photoelectric heating processes and the ionization balance of the interstellar material \citep{Lepp1988}.

Although interstellar PAH molecules are thought to originate in the winds of carbon-rich asymptotic giant branch (AGB) stars, there is little observational support for this idea. In a complete ISO/SWS survey of 50 carbon-rich AGB stars, \cite{Boersma2006} detected PAH emission in only one source, TU~Tau, a carbon-rich AGB star with a hot A2 companion providing UV radiation. Recently, \cite{Sloan2007} has detected PAH emission in an R-type carbon-rich giant with a circumstellar disk, which is the likely region where the PAH emission originates. This lack of PAH detections in carbon-rich AGB stars is most likely the result of the high optical opacities of carbon-rich dust: when formed, the PAH molecules are shielded from optical and ultraviolet photons by the opaque carbonaceous circumstellar material \citep{Jaeger1998}.

S-type stars are objects on the ascent of the AGB. They pass through the phase where the photosphere turns from oxygen-rich to carbon-rich ($\mathrm{C}/\mathrm{O} \approx 1$). Because S-type stars are not yet forming carbon-rich dust, they have less opaque circumstellar shells than their carbon-rich successors \citep{Jaeger2003}. Nevertheless, due to shock-induced non-equilibrium chemistry effects, S-type stars could display carbon-rich molecules \citep{Cherchneff2006}. Recent studies of ISO/SWS spectra of S-type AGB stars indeed show the presence of molecules like HCN and C$_2$H$_2$ \citep{Yang2007, Hony2009}.

We have studied the data obtained with the Spitzer Infrared Spectrograph \citep[IRS,][]{Houck2004} of a sample of 90 intrinsic\footnote{Intrinsic S-type stars are not enriched in C or s-process elements by accretion from an evolved companion, but through internal nucleosynthesis} S-type stars (Program ID 30737, P.I. S.\,Hony). A description of the sample can be found in \citet{Cami2009}. They present the detection of SiS absorption bands, in a large subset of the sample. In this paper, we present the detection of PAH emission around four cool S-type AGB stars. Three of the four objects extend the known correlation between the centroid wavelength of the 7.9~\um PAH feature and the stellar effective temperature to lower temperatures and more redshifted features. We argue that the hydrocarbon molecules have a high aliphatic/aromatic ratio. Since AGB stars are important producers of the dust in the interstellar medium, we might be looking at the hydrocarbon mixture, as formed in the current day wind.

\section{Observations and data reduction}
The Spitzer IRS spectra were obtained in the low-resolution staring mode, which places the star in two nod positions. This is first done in the short-low (SL) apertures, and then in two long-low (LL) apertures. The resulting spectrum covers the 5 to 37\,$\mu$m region with a typical resolution of R$\sim$100. 

For the data reduction we used the FEPS pipeline, developed for the Spitzer legacy program ``Formation and evolution of planetary systems''. A detailed description can be found in \citet{Hines2005}. For the extraction we used the intermediate droop data products as delivered by the SSC, together with the SMART reduction package tools described in detail in \citet{Higdon2004}. The spectra are background-corrected by subtracting the images of the two slit positions for each module and order. The IRS spectrum of all stars is de-reddened using the extinction maps from \citet{Schlegel1998} and the IR extinction law of \cite{Chiar2006}.

\section{The observed emission features}
The PAH molecules exhibit very characteristic emission features in the 5-14~\um region, the most prominent at 6.2, 7.9, 8.6, and 11.2 \um. These features arise from the bending and stretching of the carbon and hydrogen bonds in the large molecules.

\begin{figure}
\resizebox{\hsize}{!}{\includegraphics{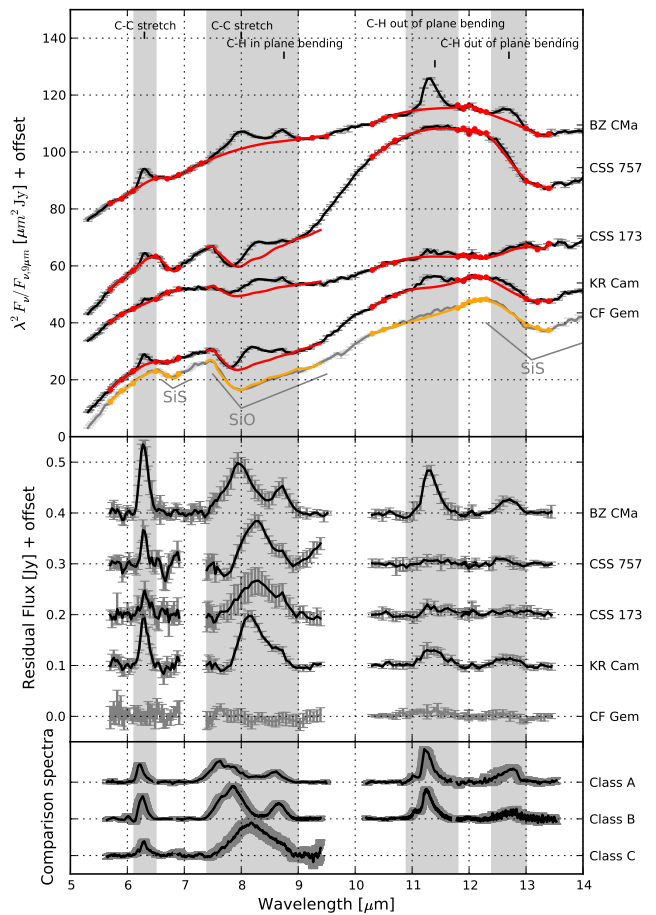}}
\caption{\emph{Top panel:} The infrared spectra of the 4 stars (black) and the continuum  estimates (red) with the characteristic infrared emission features from PAH molecules, shown in Rayleigh units. As a reference, we show the spectrum of CF~Gem, a naked photosphere, with estimated continuum. The spectrum of CSS~757 shows weak PAH emission, as well as a dust emission feature at 11~\um. \emph{Middle panel:} The residual flux. \emph{Bottom panel:} Rescaled ISO-SWS spectra of prototypes of the different PAH classes: NGC 1333-3 (Class A), HD~44179 (Class B) and CRL~2688 (Class C) \citep{Peeters2002, vanDiedenhoven2004}.}
\label{fig:allpahs}
\end{figure}

\begin{table}
\caption{Anchor points of the continuum splines.}     
\label{table:anchorpoints}                          
\centering
\begin{tabular}{ccc}
\hline\hline
Feature & \multicolumn{2}{c}{Anchor Points}        \\
        & Class B/C sources &     Class A sources  \\
\um     &     \um         &     \um                \\
\hline
6.2  & 5.7, 5.9, 6.1, 6.5, 6.7, 6.9            & idem \\
7.9  & 7.0, 7.2, 7.4, 9, 9.2, 9.4              & 6.6, 6.8, 7.0, 9.0, 9.2, 9.4 \\
11.2 & 10.3, 10.5, 10.7, 11.8, 12.0, 12.2      & idem \\ 
\hline
\end{tabular}
\end{table}

\begin{table*}
\caption{The 4 S-type stars with significant emission features at 6.2, 7.9, and 11.2 \um.}
\label{table:pahs}
\centering
\begin{tabular}{ccccccccccccc}
\hline\hline
\multicolumn{3}{c}{Stellar parameters} & 6.2~\um & \multicolumn{4}{c}{7.9~\um} & 11.2~\um \\
\hline
Name         & T$_{\mathrm{eff}}^{\mathrm{a}}$ & C/O$^{\mathrm{a}}$ & Flux$^{\mathrm{b}}$ & Class$^{\mathrm{c}}$ & $\lambda_\mathrm{peak}$ & $\lambda_\mathrm{cen}$ & Flux$^{\mathrm{b}}$ & Flux$^{\mathrm{b}}$ \\
             & K               &                 & erg\,cm$^{-2}$\,s$^{-1}$ & & \um & \um & erg\,cm$^{-2}$\,s$^{-1}$ & erg\,cm$^{-2}$\,s$^{-1}$ \\
\hline
CSS 173      & $3600 \pm 150$  & $0.97 \pm 0.01$ & $3.0 \pm 0.7 \times 10^{-13}$  & C & $8.25 \pm 0.05$ & $8.38 \pm 0.10$ & $1.6 \pm 0.2 \times 10^{-12}$ & $1.1 \pm 0.2 \times 10^{-13}$ \\
KR Cam       & $3400 \pm 150$  & $0.97 \pm 0.01$ & $6.2 \pm 0.6 \times 10^{-13}$  & C & $8.16 \pm 0.05$ & $8.31 \pm 0.06$ & $1.5 \pm 0.1 \times 10^{-12}$ & $1.3 \pm 0.2 \times 10^{-13}$ \\
BZ CMa       & $3400 \pm 150$   & $0.98 \pm 0.01$ & $20  \pm 2   \times 10^{-13}$  & B & $7.97 \pm 0.04$ & $8.02 \pm 0.04$ & $3.4 \pm 0.2 \times 10^{-12}$ & $9.8 \pm 0.5 \times 10^{-13}$ \\
CSS 757      & $3400 \pm 150$  & $0.97 \pm 0.01$ & $5.6 \pm 0.7 \times 10^{-13}$  & C & $8.27 \pm 0.05$ & $8.47 \pm 0.06$ & $2.2 \pm 0.2 \times 10^{-12}$ & $1.1 \pm 0.3 \times 10^{-13}$ \\
\hline                                          
\end{tabular}
\begin{list}{}{}
\item[$^{\mathrm{a}}$] Derived from the comparison of low-resolution spectroscopy with a dedicated grid of {\sc marcs} model atmospheres for S-type stars \citep[Plez et al. in preparation,][]{Gustafsson2008}. The low-resolution spectroscopy was obtained for the complete sample of S-type stars, with the William Hershel Telescope on La Palma for the sources in the northern sky (range: 4100\,\AA-8300\,\AA; resolution: 2\,\AA) and with the 1.9\,m telescope at the South African Astronomical Observatory for the southern targets (range: 4900\,\AA-7800\,\AA; resolution: 4\,\AA). The entire dataset will be presented in a forthcoming paper.
\item[$^{\mathrm{b}}$] Integrated line flux of the residual flux in the interval [6.12,6.5] \um, [7.4,9.0] \um, or [10.9,11.8] \um.
\item[$^{\mathrm{c}}$] According to \citet{Peeters2002}.
\end{list}\end{table*}

\subsection{Extraction of the PAH features}\label{subsec:extraction}
The S-type stars in the sample show a wide variety of infrared spectra, containing photospheric absorption bands of e.g. \co, \h2o, \sio, and \sis. The absorption bands in combination with the weak emission features make it difficult to estimate an accurate continuum. The method of extracting possible emission features combines the use of cubic splines for the 6.2 and 11.2~\um features and comparison stars for the 7.9-8.6 \um complex.

For the regions around 6.2 and 11.2~\um, the continuum flux was estimated by fitting a cubic spline (with smoothness zero) through anchor points at different positions next to the features (see Table~\ref{table:anchorpoints}). By applying random variations of the spectra within the calculated error bars and shifting the anchor points 2 spectral points outwards, we obtained a large set of continua. The spread on these continua was used to estimate the error introduced by fitting a spline to the data.

Strong water vapor absorption can mimic the presence of a 6.2~\um emission feature \citep[see for example Fig. A.23 in][]{Cami2002}. However, the water vapor bands can be easily recognized by comparing the full spectrum with synthetic spectra, and are generally weak in S-type stars. We excluded five stars in which the detection of a 6.2~\um feature was due to this effect.

BZ~CMa is the only star in our sample of S-type stars with possible PAH emission, which does not show the SiO absorption band in the region containing the 7.9 and 8.6~\um PAH emission features. For this star, we used a spline at the anchor points mentioned in Table~\ref{table:anchorpoints} to estimate the continuum. The other stars contain the SiO band head, which varies in depth throughout the sample of S-type stars. Because this absorption band is asymmetrical, a spline is not a good estimate of the photospheric continuum. Instead, we used the 40 out of 90 sample stars without dust emission (naked photospheres) as template stars, fitted in the regions 7.4-7.9~\um and 9-9.4~\um where the contribution of the PAH emission is negligible. We used the spread on the 5 best-fitting template stars as a conservative estimate of the error caused by fitting these spectra.

We calculated the integrated flux of the three continuum-subtracted features. When all features had a line flux exceeding the 3-sigma level, we considered this a positive detection of PAH emission. In Table~\ref{table:pahs} we present the 4 stars with a positive detection of PAH emission. Figure \ref{fig:allpahs} shows the spectra, the estimated continua and the residual fluxes of these stars. We include the spectrum of CF~Gem, a star without PAH emission, for comparison and to indicate the quality of the continuum estimates.

The detected PAH features in the infrared spectra can have an interstellar origin. To exclude this possibility, the background spectrum in the off-source nod (i.e. in the other order) was extracted. Since no significant features were found in the background spectra, we conclude that the PAH emission comes from the immediate environment of the star. Moreover, the feature profiles are very different from those observed in interstellar sources. From here on we focus only on these 4 stars with a positive detection of PAHs: BZ~CMa, KR~Cam, CSS~173, and CSS~757.

\subsection{Classification}
\citet{Peeters2002} shows that the C-C stretch modes at 6.2 and 7.9~\um display a wide spread in peak position.  They classify objects in three classes from A (bluest features) over B to C (reddest features). Because of the low spectral resolution and signal-to-noise of our data, we focused on the broad feature at 7.9~\um. For Class C the peak of this feature is shifted to 8.2~\um, marking a clear difference with the 7.9~\um feature of Class B PAHs and the 7.6~\um feature of Class A PAHs. In Table~\ref{table:pahs}, we show the peak wavelengths and the classes we derived for all stars. In Fig. \ref{fig:allpahs} the residual spectra of the S-type stars with PAH features, i.e. after continuum subtraction, are compared with PAH features from each of these 3 classes.

Based on the peak wavelength, BZ~CMa is the only star that shows the Class B PAHs, generally found in PNe, Herbig Ae/Be stars, and in some red supergiants \citep{Sylvester1998, Tielens2008, Verhoelst2009}. The three other stars are members of the small group of known Class C sources, together with a few post-AGB stars \citep{Peeters2002, Kraemer2006, Clio2009}, the Herbig Ae/Be source HD 135344B \citep{Sloan2005}, the intermediate-mass T Tauri star SU Aur \citep{Furlan2006}, the carbon-rich red giant HD~100764 \citep{Sloan2007}, and the oxygen-rich K giant HD~233517 \citep{Jura2006}.

\section{Correlation between T$_\mathrm{eff}$ and feature redshift}\label{section:correlation}
The centroid wavelength ($\lambda_\mathrm{cen}$) is the continuum-subtracted, flux-weighted barycenter of the feature. Recently, it has been claimed that the centroid wavelength of the 6.2, 7.9, and 11.2~\um features are shifted to longer wavelengths in targets with a lower effective stellar temperature \citep{Sloan2007, Keller2008, Boersma2008}.

\subsection{Extending the correlation towards cooler stars}
\begin{figure}
\resizebox{\hsize}{!}{\includegraphics{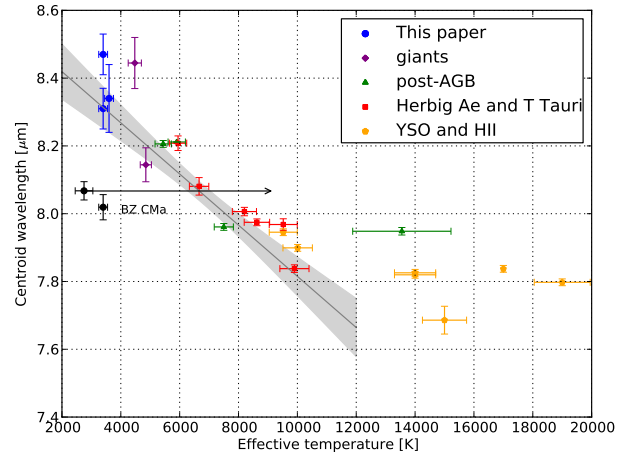}}
\caption{The centroid wavelength of the 7.9~\um feature against the stellar effective temperature. 
BZ~CMa and TU~Tau are shown as black circles. The arrow indicates the shift of TU~Tau to the temperature of the A2 companion. The gray line and area represent the correlation, quantified in Eq. (\ref{eq:corr}) with the corresponding 99\% confidence interval.}
\label{fig:sloanfig}
\end{figure}

In Fig. \ref{fig:sloanfig} we show the centroid wavelength of the 7.9~\um feature for all stars in our sample, supplemented by the centroid wavelengths of a comparison sample consisting of all Class C sources with observed 7.9~\um features \citep{Kraemer2006, Furlan2006, Jura2006, Sloan2007}, the carbon-rich AGB star TU~Tau \citep{Boersma2006}, the Herbig Ae stars without silicate emission from \citet{Keller2008}, and all PAH sources in \cite{Peeters2002} with stellar temperatures below 20\,000~K. The continuum flux was estimated by fitting a cubic spline through the anchor points given in Table~\ref{table:anchorpoints}, as described in Sect. \ref{subsec:extraction}. To calculate the centroid wavelengths, we fit a line to the left and right of the 7.9~\um feature and removed the 8.6~\um feature. Redefining the centroid wavelength for \emph{all} sources, using the same methods as for the S-type stars, allows for a quantative description of the correlation.

Our sample confirms the strong correlation between the effective temperature of the central star and the centroid position of the 7.9~\um feature. Moreover, the PAH features in the cool S-type stars are redshifted the most. The correlation can be approximated by a linear relationship for all temperatures below 12\,000~K. A linear regression analysis results in
\begin{equation}
\label{eq:corr}
 \lambda_\mathrm{cen}\ [\mu\mathrm{m}] = (8.05 \pm 0.01) 
- (7.54 \pm 0.60) \times (\mathrm{T}_\mathrm{eff}\ [10^{5}\ \mathrm{K}] - 0.077)\ \ \ \ \ 
\end{equation}
with a Pearson correlation coefficient of $R = -0.90$ and a two-tailed probability much lower than 0.1 percent, indicating a very strong correlation. All sources with higher temperatures have Peeters Class A features ($\lambda_\mathrm{cen}$ = 7.84 $\pm$ 0.06 \um). A similar correlation is present for the 6.2 and 11.2\,$\mu$m features, albeit to a lesser degree.

Two stars that do not follow this linear trend are BZ~CMa and TU~Tau. The latter, however, is a known spectroscopic binary with a hot A2-type companion (T$_\mathrm{eff} \approx 9$\,400~K) \citep{Richer1972}. If we use this temperature instead of the effective temperature of the carbon star (2\,750~K), this star falls within the scatter on the correlation. For BZ~CMa we use Eq. (\ref{eq:corr}) to predict the presence of a hot A7-type companion with a temperature of approximately 8\,100~K. Additional data is required to check this prediction.

For the three other stars there is no indication of a hot companion. The detection of PAH features in the IR spectra of these stars means that PAHs can be sufficiently excited by stars with low effective temperatures, hence soft radiation fields.

\subsection{Aliphatic and aromatic material}
\cite{Sloan2007} propose an explanation for this correlation, based on qualitatively similar shifts in laboratory measurements of PAHs. Class C PAHs are not subjected to strong UV radiation fields and thus are relatively unaltered mixes of aliphatic and aromatic hydrocarbons. Class A and B PAHs have been processed by the hard radiation of the central stars, suppressing the aliphatic component \citep[e.g.][]{Goto2002, Goto2003, Goto2007}. This agrees with the results of \citet{Pino2008}, who show that Class C sources, with a redshifted 6.2\,\um feature, have a strong a strong aliphatic emission feature at 3.4\,\um.

The three targets following the correlation at very low temperatures can be classified as \emph{extreme} Class C sources, with very broad, red emission features. CSS~757, KR~Cam and CSS~173, are thus the objects with the highest aliphatic/aromatic content ratio observed in space to date, exhibiting the composition as condensed in their winds.

\section{Discussion}
Merely that we can detect hydrocarbon features around cool stars without a hot companion is remarkable. These detections push the temperature of the radiation field exciting PAHs to even lower values than before. Since these stars do not have strong UV radiation fields, the hydrocarbon mixture must be pumped by optical photons. The theoretical model presented in \cite{Li2002} predicts that optical pumping is possible for ionized PAHs and for large PAH molecules with more than 100 carbon atoms. Later refinements of this PAH model, based on laboratory measurements, show that even near-infrared photons can excite ionized PAH molecules \citep{Mattioda2005, Mattioda2005b}. The detection of hydrocarbon emission around the S-type stars implies that the molecules are either ionized or very large, or, alternatively, that they can be excited by optical photons due to the aliphatic component.

The detection of hydrocarbon emission in combination with the SiO absorption band in the stellar photosphere points to mixed chemistry. This is unexpected if the ongoing chemistry in the circumstellar environment of S-type stars is in thermal equilibrium \citep{Ferrarotti2002}. Because the presence of SiO in the photosphere points to an oxygen-dominated chemistry, the majority of C atoms are bound in CO molecules. However, in \citet{Cherchneff2006}, the nonequilibrium chemistry in the inner wind of AGB stars is studied, and the results differ strongly from those found by \citet{Ferrarotti2002}. For stars with an equal amount of carbon and oxygen ($\mathrm{C/O} = 1$) nonequilibrium, shock-driven chemistry predicts \c2h2 molecules in the outer regions of the stellar outflow. These carbon-rich molecules are the building blocks of PAHs \citep{Cherchneff1992, Cau2002}. This is consistent with all C/O ratios found for these stars being very close to unity (see Table~\ref{table:pahs}). When compared to our entire sample of 90 S-type stars, the C/O ratios of these four stars are in the top 10\%, with the others being less carbon-rich.

\section{Conclusions}\label{sec:conclusions}
In this paper we present a positive detection and identification of PAH emission in 4 S-type AGB stars. In this small sample we see a clear difference between the strong PAH emission in BZ~CMa, which can be classified as Peeters Class B, and the much weaker PAH features seen in the other sources of Class C. We predict that BZ~CMa is a binary system with a hot, late-A-type companion.

Our data are consistent with the strong correlation found between the centroid wavelength of the 7.9~\um feature and the temperature of the central star. They extend this correlation towards lower temperatures and more redshifted features. This is consistent with the hypothesis that Class C PAHs are hydrocarbon molecules with a high aliphatic/aromatic content ratio found around stars with weak UV radiation fields. The hydrocarbons around CSS~757, KR~Cam, and CSS~173 thus represent the composition as condensed in the AGB wind, before entering the interstellar medium where harsh UV radiation alters their chemical structure.

\bibliographystyle{aa}
\bibliography{14254References.bib}

\begin{acknowledgements}
The authors wish to thank A.~G.~G.~M.~Tielens, J.~Bouwman, J.~Cami, and P.~Degroote for helpful comments and discussion of this study and the general context. K. Smolders, J. Blommaert, L. Decin, and H. Van Winckel acknowledge support from the Fund for Scientific Research of Flanders under the grant G.0470.07. S. Van Eck is an F.N.R.S Research Associate.
\end{acknowledgements}

\end{document}